\documentclass{elsart}
\usepackage{psfig,latexsym}

\begin{document}
\begin{frontmatter}

\title{Application of random matrix theory to quasiperiodic
       systems} 

\author[Chemnitz]{Michael Schreiber},
\author[Chemnitz]{Uwe Grimm},
\author[Chemnitz]{Rudolf A.\ R\"{o}mer} and 
\author[Chemnitz,China]{Jian-Xin Zhong}
\address[Chemnitz]{Institut f\"{u}r Physik, Technische
        Universit\"{a}t, D--09107 Chemnitz, Germany}
\address[China]{Department of Physics, Xiangtan University, Xiangtan
        411105, P. R. China}

\begin{abstract}
  We study statistical properties of energy spectra of a tight-binding
  model on the two-dimensional quasiperiodic Ammann-Beenker tiling.
  Taking into account the symmetries of finite approximants, we find
  that the underlying universal level-spacing distribution is given by
  the Gaussian orthogonal random matrix ensemble, and thus differs
  from the critical level-spacing distribution observed at the
  metal-insulator transition in the three-dimensional Anderson model
  of disorder. Our data allow us to see the difference to the Wigner
  surmise.
\end{abstract}
\end{frontmatter}

\begin{figure}[t]
\centerline{\psfig{figure=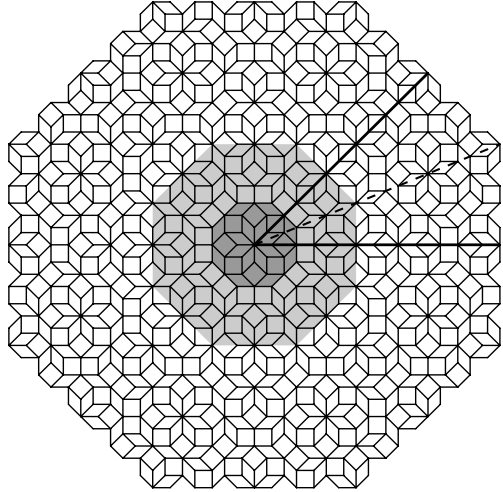,width=0.5\textwidth}}
  \caption{
    Octagonal cluster of the Ammann-Beenker tiling with $833$
    vertices and exact $D_8$ symmetry around the central vertex
    as indicated by the solid and dashed lines. Shadings indicate 
    successive inflation steps of the central octagon.}
\label{fig1}
\end{figure}

In a recent paper \cite{zgrs}, we investigated energy spectra of
quasiperiodic tight-binding models, concentrating on the case of the
octagonal Ammann-Beenker tiling \cite{AB} shown in Fig.~\ref{fig1}.
The Hamiltonian is restricted to constant hopping matrix elements
along the edges of the tiles in Fig.~1.  Previous studies of the same
model had led to diverging results on the level statistics: For
periodic approximants, level repulsion was observed \cite{BS,PJ}, and
the level-spacing distribution $P(s)$ was argued to follow a
log-normal distribution \cite{PJ}. On the other hand, for octagonal
patches with an exact eightfold symmetry and free boundary conditions,
level clustering was found \cite{ZY}. On the basis of our numerical
results for $P(s)$ and the spectral rigidity $\Delta_3$ \cite{MH},
compiled in Ref.~\cite{zgrs}, we concluded that the underlying
universal level-spacing distribution of this system is given by the
Gaussion orthogonal random matrix ensemble (GOE)
\cite{MH,WD}. Concerning the contradictory results of previous
investigations, we attribute these to the non-trivial symmetry
properties of the octagonal tiling. The periodic approximants studied
in Refs.~\cite{BS,PJ} show, besides an exact reflection symmetry, an
``almost symmetry'' under rotation by 90 degrees which may influence
the level statistics \cite{MH}, whereas the octagonal patches used in
Ref.~\cite{ZY} possess the full $D_8$-symmetry of the regular octagon.
Hence the level statistics observed in this case is that of a
superposition of seven completely independent subspectra, and
therefore rather close to a Poisson law.

To arrive at this conclusion, we considered in Ref.~\cite{zgrs}
different patches that approximate the infinite quasiperiodic tiling,
both with free and periodic boundary conditions. Exact symmetries were
either exploited to block-diagonalize the Hamiltonian, thus splitting
the spectrum into its irreducible parts, or avoided altogether by
choosing patches without any symmetries. 

\begin{figure}[b]
\centerline{\psfig{figure=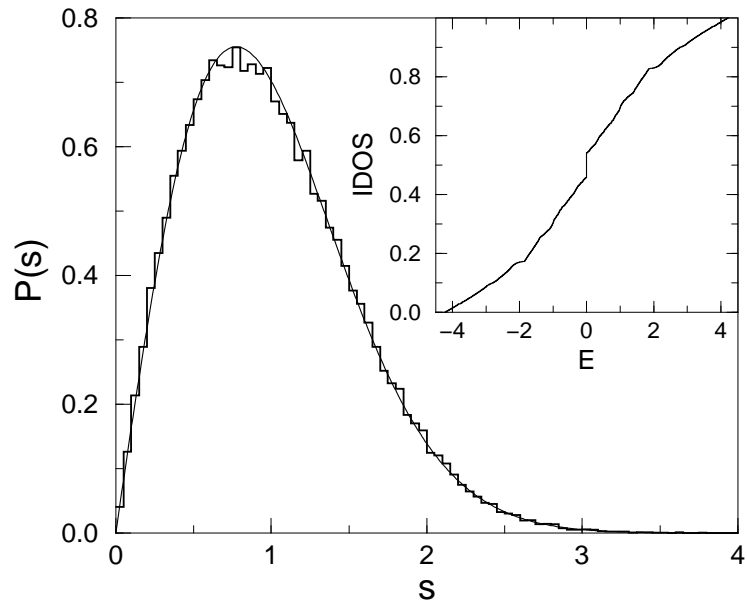,width=0.7\textwidth}}
  \caption{IDOS (inset) for the $D_8$-symmetric patch with
    $N=157369$ vertices, and $P(s)$ averaged over the three largest
    sectors. The smooth line denotes $P_{\rm GOE}(s)$.}
\label{fig2}
\end{figure}

Here, we concentrate on the $D_8$-symmetric octagonal patch shown in
Fig.~\ref{fig1}. For this case, the Hamiltonian matrix splits into ten
blocks according to the irreducible representations of the dihedral
group $D_8$, resulting in seven different independent subspectra as
there are three pairs of identical spectra. In Fig.~\ref{fig2}, we
show the integrated density of states (IDOS) for a patch, which
contains $N=157369$ vertices and corresponds to three more inflation
steps performed on the patch of Fig.~\ref{fig1}. Apparently, the IDOS
is rather smooth, and the only prominent feature that shows up, apart
from a few small gaps, is the huge fraction (13077 of 157369, hence
about 8.3\%) of exactly degenerate eigenvalues in the band center. For
the level-spacing distribution $P(s)$, these do not matter as they
would only contribute to $P(0)$, wherefore we can neglect them
completely. The IDOS shown in Fig.~\ref{fig2} is fitted to a cubic
spline which is then used to ``unfold'' the spectrum \cite{HS}, i.e.,
to correct for the non-constant density of states, what is necessary
if we want to compare to results of random matrix theory.

\begin{figure}[t]
\centerline{\psfig{figure=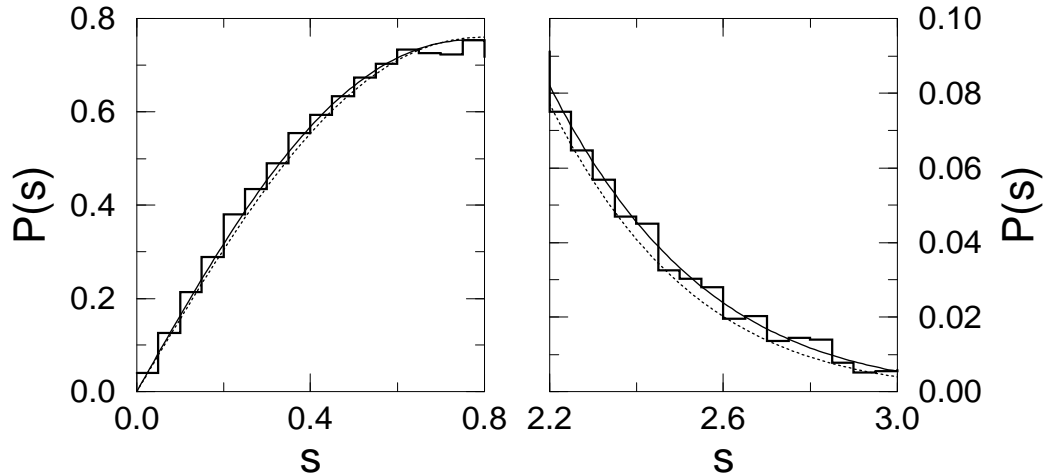,width=\textwidth}}
  \caption{Small- and large-$s$ behaviour of $P(s)$ of
  Fig.~\protect\ref{fig2}, compared to $P_{\rm GOE}(s)$ (solid line)
  and $P_{\rm W}(s)$ (dashed line).}
\label{fig3}
\end{figure}

The level-spacing distribution $P(s)$ for the unfolded spectra is
shown in Figs.~\ref{fig2} and \ref{fig3}, measured in units of the
mean level spacing. Here, we averaged over the three largest
subspectra, each of which contains $18043$ levels after removing the
degenerate states in the band center. The resulting histogram is
compared to the GOE distribution $P_{\rm GOE}(s)$ (solid lines in
Figs.~\ref{fig2} and \ref{fig3}) and, focusing on the small- and
large-$s$ behaviour, also to the Wigner surmise $P_{\rm W}(s)$ (dashed
lines in Fig.~\ref{fig3}). Apparently, the level-spacing distribution
of the quasiperiodic Hamiltonian is well described by random matrix
theory, and one can clearly see that $P_{\rm GOE}(s)$ fits the
numerical data even better than $P_{\rm W}(s)$.

\begin{figure}[t]
\centerline{\psfig{figure=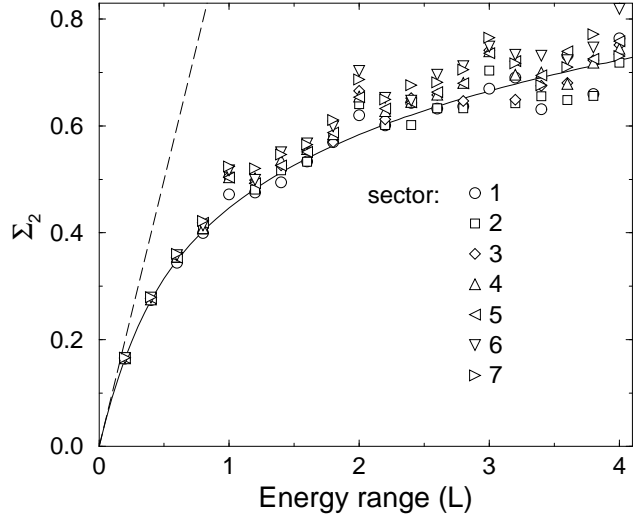,width=0.6\textwidth}}
  \caption{$\Sigma _2$ statistics for the seven independent subspectra of 
    the $D_8$-symmetric octagonal patch with $157369$ vertices. The
    lines indicate the GOE (solid) and Poisson (dashed) behaviour.}
\label{fig4}
\end{figure}

Fig.~\ref{fig4} shows the corresponding $\Sigma_2$ statistics
\cite{MH}, compared to the exact GOE result. The $\Sigma_2$ statistics
measures the fluctuation in the number of energy levels $n$ in an
energy range $L$, i.e., $\Sigma_2=\langle n^2\rangle-\langle
n\rangle^2$ where $\langle . \rangle$ denotes the spectral average.
Again, the agreement with our numerical results is good, supporting
the conclusion that the underlying universal level statistics is
described by the GOE.  Because typical eigenstates in our model are
expected to be multifractal, one might have expected that one finds a
``critical'' level-spacing distribution as observed at the
metal-insulator transition in the three-dimensional Anderson model of
disorder \cite{ZK} --- however, this is clearly not the case.

\begin{thebibliography}{9}

\bibitem{zgrs}
J.~X.~Zhong, U.~Grimm, R.~A.~R\"{o}mer, and M.~Schreiber, 
Phys. Rev. Lett. {\bf 80}, 3996 (1998).

\bibitem{AB}
R.\ Ammann, B.\ Gr\"{u}nbaum, and G.\ C.\ Shephard,
Discrete Comput.\ Geom.\ {\bf 8}, 1 (1992);
M.\ Duneau, R.\ Mosseri, and C.\ Oguey, J.\ Phys.\ A {\bf 22}, 4549 (1989).

\bibitem{BS}
V.~G.~Benza and C.~Sire,
 Phys.\ Rev.\ B {\bf 44}, 10343 (1991).

\bibitem{PJ}
F.~Pi\'{e}chon and A.~Jagannathan,
 Phys.\ Rev.\ B {\bf 51}, 179 (1995).

\bibitem{ZY}
J.~X.~Zhong and H.~Q.~Yuan,
in {\em Quasicrystals: Proceedings of the 6th Int.\
  Conference}, eds.\ S.~Takeuchi and T.~Fujiwara (World Scientific,
Singapore, 1998).

\bibitem{MH}
M.~L.~Mehta, 
{\em Random Matrices}, 2nd ed.\
(Academic Press, Boston, 1990);
F.~Haake, 
{\em Quantum Signatures of Chaos}, 2nd ed.\
(Springer, Berlin, 1992).

\bibitem{WD}
E.~P.~Wigner, 
Proc.\ Cambridge Philos.\ Soc.\ {\bf 47}, 790 (1951);
F.~J.~Dyson,
J.\ Math.\ Phys.\ {\bf 3}, 140 (1962).

\bibitem{HS}
E.~Hofstetter and M.~Schreiber,
Phys.\ Rev.\ B {\bf 48}, 16979 (1993); 
{\bf 49}, 14726 (1994);
Phys.\ Rev.\ Lett.\ {\bf 73}, 3137 (1994).

\bibitem{ZK}
I.~K.~Zharekeshev and B.~Kramer,
Phys.\ Rev.\ Lett.\ {\bf 79}, 717 (1997).

\end{thebibliography}
\end{document}